\documentclass[runningheads]{llncs}
\usepackage{graphicx}
\usepackage{multirow}
\usepackage[vlined,linesnumbered,ruled]{algorithm2e}
\usepackage{siunitx}
\usepackage{subcaption}
\usepackage[most]{tcolorbox}
\usepackage[export]{adjustbox}

\begin{document}
\title{Code Coverage Aware Test Generation Using Constraint Solver}
%
%\titlerunning{Abbreviated paper title}
% If the paper title is too long for the running head, you can set
% an abbreviated paper title here
%
% \author{Krystof Sykora\inst{1}\orcidID{0000-1111-2222-3333} \and
% Bestoun S. Ahmed\inst{2,3}\orcidID{1111-2222-3333-4444} \and
% Miroslav Bures\inst{3}\orcidID{2222--3333-4444-5555}}

\author{Krystof Sykora\inst{1} \and
Bestoun S. Ahmed\inst{2} \and
Miroslav Bures\inst{1}}

%
%\authorrunning{F. Author et al.}
% First names are abbreviated in the running head.
% If there are more than two authors, 'et al.' is used.
%
\institute{Department of Computer Science, Faculty of Electrical Eng, Czech Technical University, Prague, Czech Republic \\ \email{\{sykorkry,buresm3\}@fel.cvut.cz} \and
Department of Mathematics and Computer Science, Karlstad University, Sweden
\email{bestoun@kau.se}}
\maketitle              % typeset the header of the contribution
\begin{abstract}
Code coverage has been used in the software testing context mostly as a metric to assess a generated test suite's quality. Recently, code coverage analysis is used as a white-box testing technique for test optimization. Most of the research activities focus on using code coverage for test prioritization and selection within automated testing strategies. Less effort has been paid in the literature to use code coverage for test generation. This paper introduces a new Code Coverage-based Test Case Generation (CCTG) concept that changes the current practices by utilizing the code coverage analysis in the test generation process. CCTG uses the code coverage data to calculate the input parameters' impact for a constraint solver to automate the generation of effective test suites. We applied this approach to a few real-world case studies. The results showed that the new test generation approach could generate effective test cases and detect new faults.

\end{abstract}

\section{Introduction}

In software engineering, regression testing has become a common practice to be used during test development. In this practice, testers begin by rerunning existing test suites to validate new software-under-test (SUT) functionality. However, this approach faces many problems as existing tests decrease their ability to detect SUT faults. This paper shows how Test suite augmentation techniques can be used to solve this problem\footnote{This paper is accepted for publication in the 2nd International Workshop on Automated and verifiable Software sYstem DEvelopment
}. 

To improve the test generation, tools like code coverage analysis and constraint solving are commonly utilized. We use these techniques in the Code Coverage-based Test Case Generation (CCTG) method. This approach is akin to Test suite augmentation techniques \cite{test_suite_augmentation}, which is commonly used in regression testing. Augmentation is employed to adjust existing test cases by analyzing changes in the SUT. Practitioners of Test suite augmentation techniques believe the use of preexisting test cases can improve that test suites \cite{directed_test_suite_augmantation}. To promote these test cases, the SUT is analyzed using code coverage and other criteria to prioritize how changes to the test suites should be conducted. In our method, we focus on the second part of the augmentation approach -- code coverage. By analyzing the code coverage of preexisting generated test cases, we can determine test parameters' impact. This information is then used for the test generation.

CCTG can be used for the test suite augmentation method by utilizing the code coverage data and test models to refine the test generation process. To generate the test cases, CCTG first creates the test data sets used for the input interaction. A test model consists of the SUT parameters classes (i.e., possible parameter values). The results are sets of specific SUT parameter values, which will be referred to as test cases. The parameter classes are then used to generate random test cases to be executed for code coverage monitoring. This process generates coverage data related to each generated test case. If a specific test parameter change affects the coverage consistently or significantly, the parameter's weight also increases. The weight indicates the extent to which each parameter is used and permuted when generating a new set of test cases. The process as a whole contributes towards the test suite augmentation. Here, the code coverage data used for the augmentation may also minimize the number of test cases and generate more effective test cases in terms of fault-finding.

As in most modern software systems, the test suite augmentation also suffers from input interaction constraint problems \cite{constrained_interaction_testing_summ}. Here, we deployed a constraint solver within our approach during the test generation process to resolve the input values' constraints. The solver makes sure no test cases are generated with meaningless interactions with the SUT. The resulting test cases should resemble the general workflow of test cases \cite{automated_generation_for_workflows} as the code coverage analysis motivates variety in decision paths, and the constraint solver ensures that the combinations remain reasonable.

\section{Background}\label{Background}

Classically, code coverage has been used as an analytical approach with the test suite execution. The approach is also used with the test suite generation strategies to maximizing the code-base covered by generating more test cases \cite{ieee_automatic_generation_worst_case_input}. We have recently used more advanced techniques function as gray-box methods for test case analysis and test generation \cite{code_aware_combinatorial_interaction_testing}. We considered the code's internal structure while augmenting a generated test suite for Combinatorial Interaction Testing (CIT). This approach improved the test generation process by studying the program's code to determine individual parameters' impacts. We used this impact factor to select the input parameters and values for the test cases. In this regard, the CCTG strategy proposed in this paper is very similar. However, examining the internal code structure can prove costly, in terms of a manual analysis of the input parameters and values. So, the CCTG simplifies this process by automatically evaluating the parameter impact, requiring only a test model of parameters and their possible values. It can also explore various hidden criteria for test generation (such as SUT configuration and state).

A common problem for test generation is that the test models don't consider specific constraints that would make the test cases useless or nonsensical. To avoid this, test generation methods \cite{Gargantini2017,Jian2012} rely on constraint solvers to eliminate such undesirable combinations. Such constraint or rather Boolean satisfiability problem-solvers \cite{BALINT2015120} (SAT solvers) are also employed in the CCTG method. The solver's additional benefit is that the CCTG user may incorporate the SAT constraints into the test model to prevent the generation of unwanted test cases and correct the focus of the test cases. If new functionality is added to the SUT, it can be specified that some parameters (representing the choice to use the new functionality) must be used. This allows the tester to influence the test generation by only adjusting the test model without changing the code coverage analysis results.

While the CCTG finds relations to the previously mentioned techniques, it is primarily a test suite augmentation method \cite{test_suite_augmentation}. Our approach is based on an innovative idea of augmenting test cases using the coverage criteria. The CCTG method, however, uses randomly generated test cases to establish the coverage data. This data is then used to generate what is essentially the first set of actual test cases.

\section{The proposed method}\label{Method}
In this section, we examine each step of the CCTG method. The following subsections illustrate these steps in detail. 

\subsection{Determining parameter weight}

The initial step in the CCTG methods is the code coverage analysis. This process is used to calculate the impact (weight) of SUT parameters that will be used for test generation.

\subsubsection{SUT Test model}

For each SUT, a model (shown in figure \ref{fig:ModelAnalysis}(a)) for interaction consisting of the input parameters and constraints among them. The parameters are represented as the P[n] array, where each index P[1], P[2],... P[n] represents one of the test model parameters. Each parameter has a value. The value V is the possibility of a parameter. This can be represented in two ways, either the V is an array of the possible values for parameters (booleans or enums) from the SUTs perspective, or the value can be represented as a number with a range. In the case of range representation, the parameter depth [d] is added to represent the number of values chosen from the range for further test generation. The other parameter property in the model is its weight, which is graded as a float ranging from 0 to 1.

\begin{figure}
    \centering
    \begin{subfigure}{.4\linewidth}
        \includegraphics[scale=0.5]{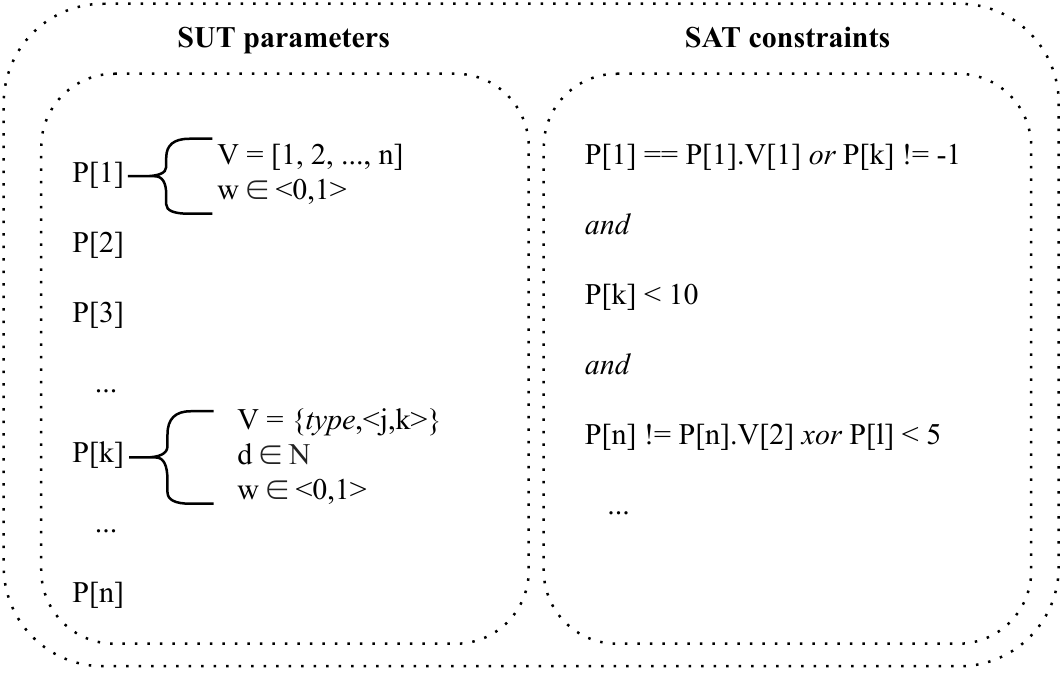}
        \caption{SUT test model}
    \end{subfigure}
    \hskip2em
    \begin{subfigure}{.4\linewidth}
        \includegraphics[scale=0.5]{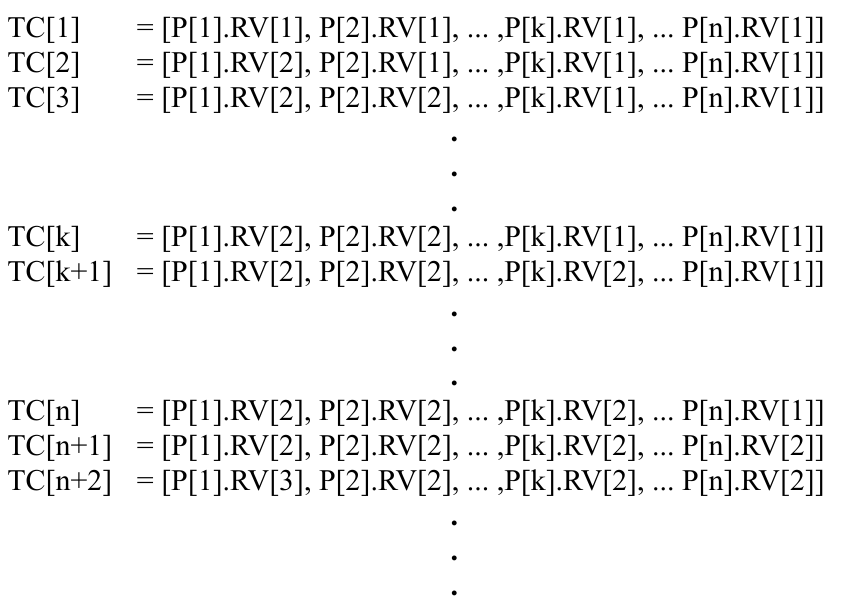}
        \caption{Test case analysis}
    \end{subfigure}
    \caption{SUT modeling and parameter analysis}
\end{figure}

The second part of the model is a set of SAT constraints to ensure that the conflicting decision is not selected. This model is used to generate the initial test cases for code coverage analysis and generate the resulting test set for the actual testing. For regression testing, the first set of random test cases can be replaced with a set of regression tests from earlier stages of the SUT development.

\subsubsection{Test case analysis and generation}

\begin{figure}
\centering
\includegraphics[width= 2.5 in]					{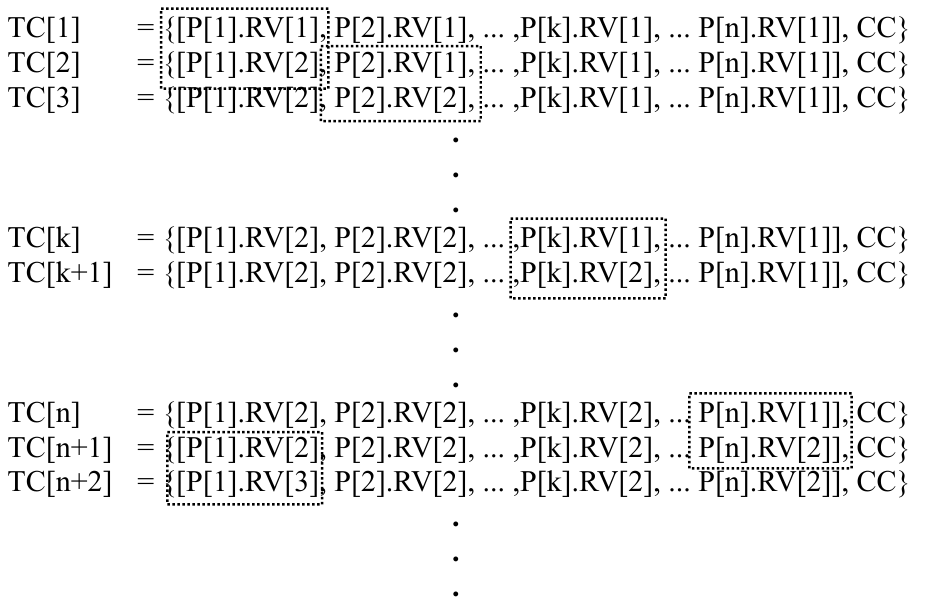}
\caption{Test case selection}
\label{fig:test_case_selection}
\end{figure}

In contrast to other strategies, maximizing of code coverage is not a direct goal for CCTG. The strategy relies on using the coverage data in a more specific manner more targeted on condition or coverage principles. We follow the standard definition of code coverage for lines of code per individual test case. The gcov\footnote{https determines the code coverage://linux.die.net/man/1/gcov} tool and stored for later analysis for each test case. The test suite used for coverage determination is designed in a very specific way. As the initial step of generation, we select a test depth level. This level represents how many values will be selected for each parameter. The same number of values as overall test depth is selected for each parameter P and randomly ordered in an array RV. The first test case TC uses each parameter's first value in all randomly ordered lists. For every test case after that, one parameter changes the value to the next in the randomised list (P[1].RV[1] to P[1].RV[2]) as shown in \ref{fig:ModelAnalysis}(b). We proceed this way until all the arrays are exhausted. The rationale behind this approach is its effects on code coverage change that determines the parameter impact. We always select two test cases where all but one parameter have the same values to measure this. This will be explained in the Parameter weight calculation section and the figure \ref{fig:test_case_selection}

\subsubsection{Parameter weight calculation}

The initial step in this phase is to determine each parameter's effect on code coverage (i.e., the parameter weight). Using the initial test cases, the code coverage is gathered automatically. To determine the information about a specific parameter, we must take a look at a set of test cases, where the examined parameter changes, while the remaining parameters stay the same, as shown in figure \ref{fig:test_case_selection}.

Initially, a parameter is selected by the algorithm for analysis. Then, similar pair test cases are selected, except for the value of the parameter under investigation. The difference in the code-coverage results of the test cases is recorded. These steps are repeated for all remaining pairs of test cases matching the criteria of all parameters being the same, except for the changing parameter under analysis. From the derived differences in coverage, an average is calculated. This average represents an absolute value of the parameters weight. These steps are illustrated in detail in the Algorithm \ref{code:2}.

\begin{algorithm}
\caption{Steps in the code coverage analysis to determine parameter's impact}\label{code:2}
    \scriptsize
    \SetKwInOut{Input}{Input}
    \SetKwInOut{Output}{Output}
    \SetKwProg{CodeCoverageAnalysis}{CodeCoverageAnalysis}{}{}
    \CodeCoverageAnalysis{$(T_CL,P)$}{
        \Input{Test case list $T_CL$, Parameter-under-test $P$}
        \Output{Measure of the impact of parameter P on code coverage $avrg(DL)$}
        $TL$ = TestCaseList\;
        $DL$ = list of float values\;
        \While{$T_CL$ not empty}{
              $TL$ add($T_CL$ pop first item)\;
              \ForEach{$T_C$ in $T_CL$}{
                \If{each i/{$PI$} : $PV$[i] in $TC$ == $PV$[i] in $TL$[0]}{
                    $TL$ add(pop $TC$ from $T_CL$)\;
                }
            $DL$ add($max($TL$)$ – $min($TL$)$)\;
            $TL$ clear\;
            }
        }
        \KwRet{$avrg($DL$)$}\;
    }

\end{algorithm}

Maximum code coverage for the generated tests is not the aim here. Instead, by determining the impact on individual parameters' coverage, parameters are selected for permutation in test cases if they influence code coverage more.

\subsection{Test case generation}

Test cases are generated using the data of parameter weight. All unary parameters have only two possible permutations (0,1) as they can only be or not be included upon program execution. The number of binary parameter permutation is equal to 1 + the number of pre-selected possible values. The unary parameter is a set with two elements (one being the parameters exclusion) and the binary a set with 1 + (number of values) elements. All test cases created are solved for constraints.

The code-coverage based method is relying on the selected parameters to permute based on their weight. All weights are places on an axis, spanning the range from 0 to 100. Thanks to normalizing the values, their sum is 100 exactly and, as such, fills up the entire axis. For example, take 3 parameters with respective weights 15, 60, and 25. The parameters would assume the following ranges on the axis. P1 (0,15), P2 (15,75) and P3(75,100).

The first constructed test case has all parameters with their default values. A random real number is generated in the range of (0,100) to construct the next test case. A parameter whose range corresponds (on the axis) to the number generated is selected for the permutation. The selected parameter's value is permuted to the next value in the sum that represents it. The test case is then saved if it is not identical to a previously created one. The test case generation algorithm is included in Algorithm \ref{code:3}.

\begin{algorithm}
\caption{Code-coverage-based test-case generation}\label{code:3}
\scriptsize
    \SetKwInOut{Input}{Input}
    \SetKwInOut{Output}{Output}
    \SetKwProg{CVBasedTCGeneration}{CVBasedTCGeneration}{}{}
    \CVBasedTCGeneration{$(PL,n)$}{
        \Input{ParameterList $PL$, Number of test cases to be generated $n$}
        \Output{TestCaseList $T_CL$}
        $T_CL$ = list of TestCases \;
        $CT_C$ = TestCase where each $P_V$ in $P_VL$ is 0\;
        \While{$n>0$}{
            $r$ = random float value\;
            $t$ = 0\;
            \ForEach{$P$ in $PL$}{
                $t$ = $t$ + $PW$ in $P$\;
                \If{$r<=t$}{
                    permute($P$)\;
                }                
            }
            add(PL)\;
              $n$--\;
        }
        \KwRet{$T_CL$}\;
    }

\end{algorithm}

We have used the Z3 solver with this test generation to resolve the constraints. We have also developed an interface for the tool so that the relevant constraints for test cases exclusion can be imported from the test model.

\section{Experimental evaluation}\label{experiment}

To evaluate the CCTG method, we have conducted three case studies. The experiments were designed according to the Mutation Testing \cite{automatic-generation-mutation-testing} approach. For each case study, several mutants were created using a fault seeding framework. To test the CCTG method's effectiveness, the generated test cases were used for both the original version of the SUT and the mutated (faulty) version. The outputs of the two SUTs were then compared. When the outputs differ, the fault is considered to be detected. The mull \cite{mull-it-over} LLVM-based tool was used.

For the case studies, the Unix utilities Flex, grep, gzip were selected. The Software-artifact Infrastructure Repository\footnote{http provided the programs://sir.unl.edu} obtained from the Gnu site. These utilities were chosen, as the test cases generated represent their command-line arguments. Therefore, each test case is essentially a parametrized call of these utilities that produces a standard output test.

The test case structure reflects the archetype of the SUTs from the case study, standard C utilities, requiring only a set of arguments. For the case study, CCTG test cases were therefore represented by a set of command-line arguments only. The type of arguments or parameters used can be divided into a few basic categories. There is a unary argument -- which always represents a Boolean value. As the parameters are for typical command-line programs, these usually specify some functionality (e.g., --printToCommandLine) that is enabled or a specific instruction (e.g., --help). The second type is binary arguments, consisting of parameter and value (e.g., --input /inputs/file1).

For each SUT, a set of all possible parameters is gathered. This is a simple list of all unary parameters accompanied by possible values for the binary ones. In many cases, the parameter values are subjective to the testing environments. Input files are part of prepared testing inputs for a specific program. Values with range such as integers are also limited to discrete and finite selection. The range's selecting values are done either based on other testing information, chosen randomly, or at regular intervals. The selection of specific values from a range does not directly affect the experiments as the selections are final for all test generation methods. Seeded versions of SUT are executed using test cases generated by various methods to measure the test effectiveness. The methods used are compared to the code-coverage method.

As a reference, we use two methods of test generation: Random generation and unweighted method. The random generation method is based on random parameter permutation. The initial setup is very similar to the CCTG method, as prepared sets represent the parameters. Unlike the CCTG method, the random method does not rely on coverage information but approaches the selection of parameters for permutation entirely randomly. The selection of parameters for permutation and the value to which it should be permuted is determined randomly. First, a random integer in the range representing all the test case parameters is generated to select a permutation parameter. Secondly, a second random integer is generated within the range of all said parameter values minus one (the one being the previous parameter value - so as not to repeat the same test case).

\begin{figure*}
    \centering
    \begin{subfigure}[b]{0.3\textwidth}
        \includegraphics[width=\textwidth]                    {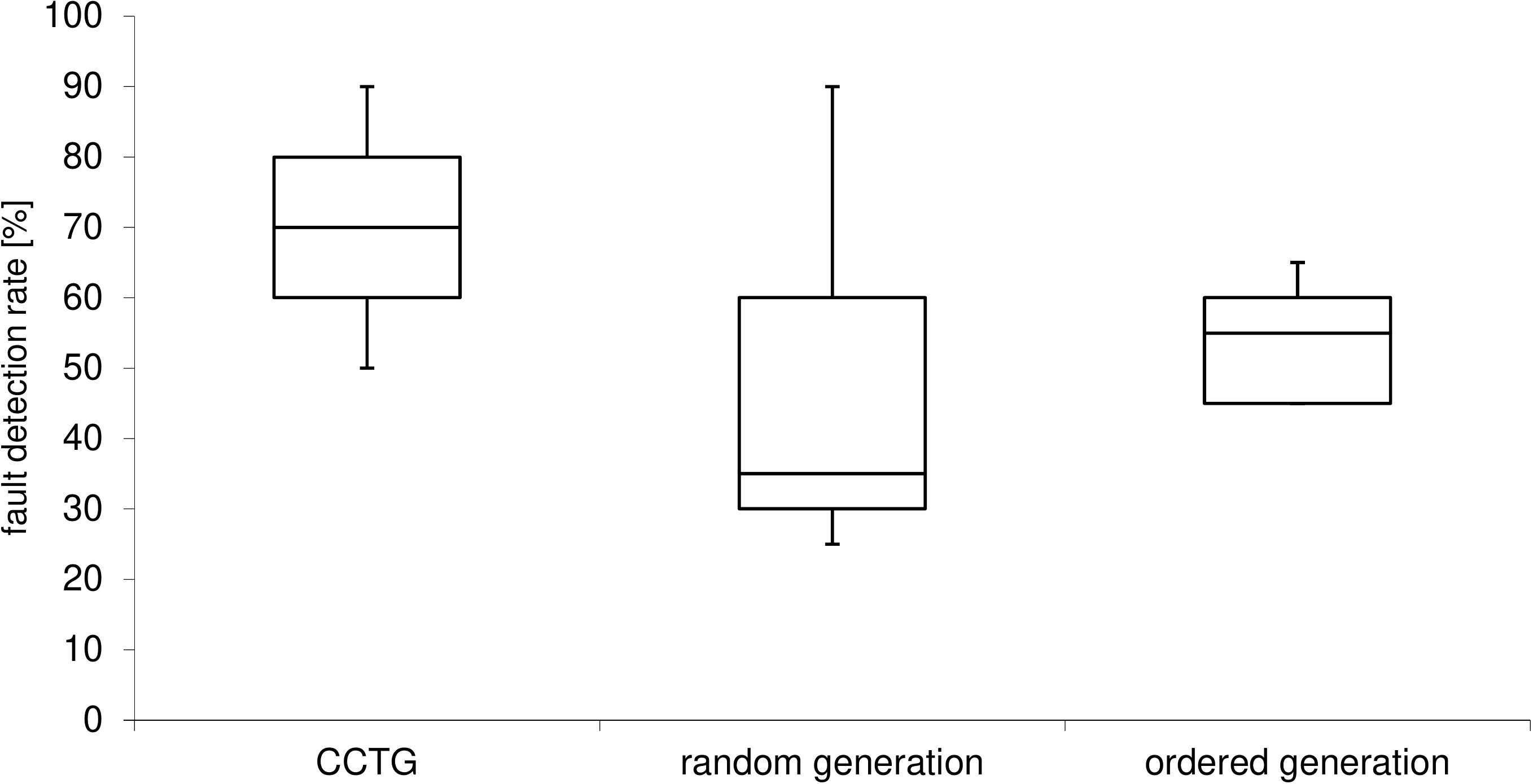}
                \vspace{-0.10cm}
        \caption{Flex \% of faults found}
        \label{fig:flex_croped}
    \end{subfigure}%
         \hfill
    \begin{subfigure}[b]{0.3\textwidth}
        \includegraphics[width=\textwidth]                    {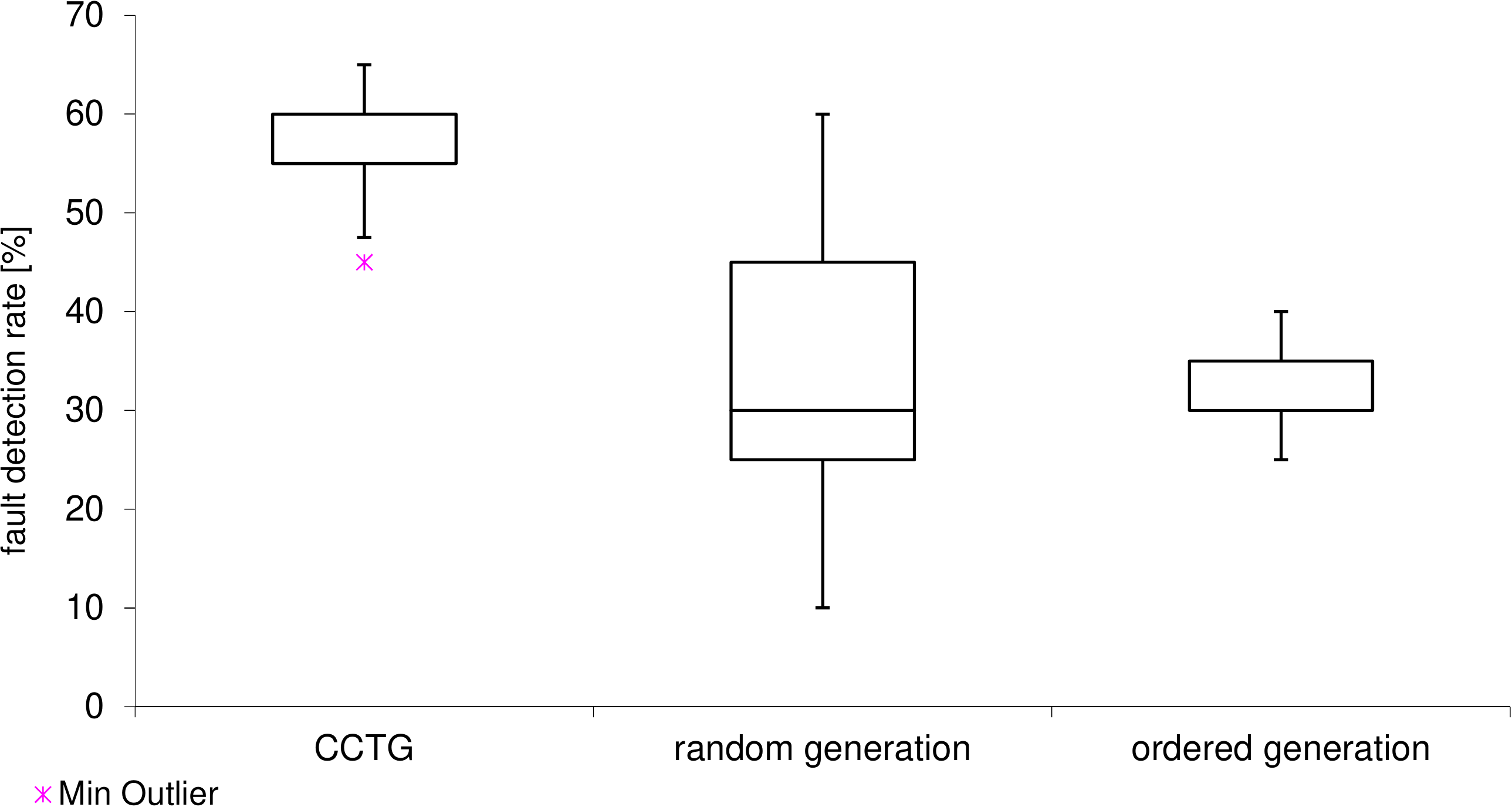}
                \vspace{-0.10cm}
        \caption{Grep \% of faults found}
        \label{fig:grep_croped}
          \end{subfigure}
          \hfill
    \begin{subfigure}[b]{0.3\textwidth}
        \includegraphics[width=\textwidth]                    {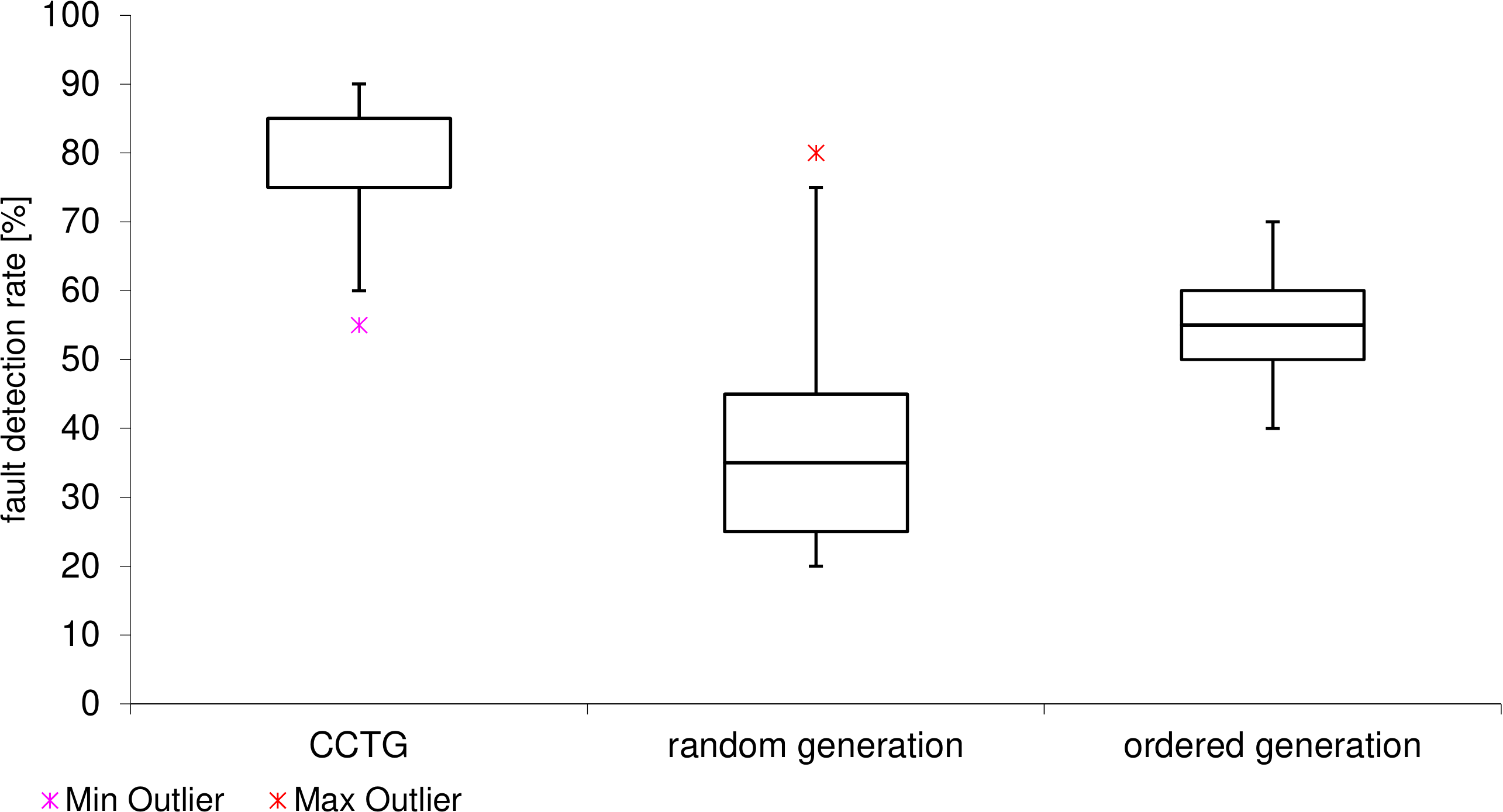}
                \vspace{-0.10cm}
        \caption{GZip \% of faults found}
        \label{fig:gzip_croped}
          \end{subfigure}
\caption{CCTG evaluation results in Box Plot}\label{fig:AllBoxes}
\end{figure*}

In each study, a set of 100 test cases is generated. This is done for all three methods and is used on 20 different seeded faults. The entire process is then repeated five times by generating a new set of test cases. The test cases were generated multiple times to account for the random elements in their generation. The code-coverage based method generally has the best results. For illustration, the results of each program are show in box-plot Figures \ref{fig:flex_croped}, \ref{fig:grep_croped}, and \ref{fig:gzip_croped}.

The Flex results are shown in Figure \ref{fig:flex_croped} and have the narrowest range for the code-coverage method – the shortest box plot. This indicates that the coverage based test cases are of very similar quality for flex. The code coverage method also has an overall higher median. However, the random method shows a tall box-plot, reflecting a completely random approach to test generation.

Figure \ref{fig:grep_croped} for the Grep experiments has all box plots of similar size. This reflects the random elements in all test generation methods. However, it does not produce such dissimilar sizes as in the Flex experiments. The overall dispersion, while similar in size, is marginally more successful for the code coverage method.

Gzip experiment results are shown in Figure \ref{fig:gzip_croped}. Here the smallest box plot represents the systematic method. While this does not correspond with results from other tests, it is not necessarily a surprise, as the systematic method has the lowest random factor of generation. The code coverage method again holds a marginally better median then the remaining two methods.

In all three cases, the medians are lower compared to the code coverage method. Most distributions in all figures are also not widely dissimilar, indicating even effectively test generating methods. It also shows the code coverage method as the most effective one.

\section{Conclusion}\label{conclusion}
This paper presented a new automated test case generation method based on the code coverage measure. The method's goal is to achieve automated test generation using the code coverage, which would also show improved performance at fault detection. Three case studies were implemented that compared out method against two other trivial approaches for test case generation. The results showed an overall improvement in the fault detection rate. The future goal is to work with a wider variety of parameters.

\section*{Acknowledgement}

This research is conducted as a part of the project TACR TH02010296 Quality Assurance System for the Internet of Things Technology. The authors acknowledge the support of the OP VVV funded project CZ.02.1.01/0.0/0.0/16\_019/0000765 “Research Center for Informatics.”

%
% BibTeX users should specify bibliography style 'splncs04'.
% References will then be sorted and formatted in the correct style.
%
 \bibliographystyle{splncs04}
 \bibliography{sample-base}
%
% \begin{thebibliography}{8}
% \bibitem{ref_article1}
% Author, F.: Article title. Journal \textbf{2}(5), 99--110 (2016)

% \bibitem{ref_lncs1}
% Author, F., Author, S.: Title of a proceedings paper. In: Editor,
% F., Editor, S. (eds.) CONFERENCE 2016, LNCS, vol. 9999, pp. 1--13.
% Springer, Heidelberg (2016). \doi{10.10007/1234567890}

% \bibitem{ref_book1}
% Author, F., Author, S., Author, T.: Book title. 2nd edn. Publisher,
% Location (1999)

% \bibitem{ref_proc1}
% Author, A.-B.: Contribution title. In: 9th International Proceedings
% on Proceedings, pp. 1--2. Publisher, Location (2010)

% \bibitem{ref_url1}
% LNCS Homepage, \url{http://www.springer.com/lncs}. Last accessed 4
% Oct 2017
% \end{thebibliography}
\end{document}